\documentstyle[prl,aps]{revtex}
\begin{document}
\draft
\title{Chemical fluctuations \\
in high-energy nuclear collisions
}
\author{Stanis\l aw Mr\' owczy\' nski\footnote{Electronic address:
mrow@fuw.edu.pl}}

\address{So\l tan Institute for Nuclear Studies, \\
ul. Ho\.za 69, PL - 00-681 Warsaw, Poland \\
and Institute of Physics, Pedagogical University, \\
ul. Konopnickiej 15, PL - 25-406 Kielce, Poland}

\date{26-th January 1999, revised 26-th April 1999}

\maketitle

\begin{abstract}

Fluctuations of the chemical composition of the hadronic system produced
in nuclear collisions are discussed using the $\Phi-$measure which has been
earlier applied to study the transverse momentum fluctuations. The measure 
is expressed through the moments of the multiplicity distribution and then 
the properties of $\Phi$ are discussed within a few models of multiparticle 
production. A special attention is paid to the fluctuations in the equilibrium 
ideal quantum gas. The system of kaons and pions, which is particularly 
interesting from the experimental point of view, is discussed in detail.

\end{abstract}

\vspace{0.5cm}
 

\vspace{0.5cm}

There are many sorts of hadrons produced in high energy collisions.
The ratios of multiplicities of particles of given species to the 
total particle number characterize the chemical composition of the 
collision final state. The composition is expected to reflect the 
collisions dynamics. Generation of the quark-gluon plasma in 
heavy-ion collisions was argued long ago \cite{Koc86} to
enhance the strange particle production. While the significant 
strangeness yield enhancement has been experimentally 
observed in the central nucleus-nucleus collisions at CERN SPS 
(see the data compilation \cite{Gaz96} and the recent review 
\cite{Ody98}), it is a matter of hot debate whether the observation 
can be indeed treated as a plasma signal. The strange hadron 
abundance is naturally described within the models assuming 
the plasma occurrence at the early collision stage (see e.g. 
\cite{Raf96,Gaz97a}) but the models, which neglect such a possibility 
(see e.g. \cite{Bra96,Cap96} or the review \cite{Sor98}), can be also 
tuned to agree with the experimental data. Thus, it would be desirable 
to go beyond the average particle numbers and see whether the strangeness 
enhancement in the central heavy-ion collisions is accompanied with 
the qualitative change of the strangeness yield fluctuations. The 
equilibrium quark-gluon plasma scenario is obviously expected to 
lead to the smaller fluctuations than the nonequilibrium cascade-like 
hadron models but the specific calculations are needed to quantify such 
a prediction. Anyhow, it seems to be really interesting to study the 
strangeness yield fluctuations on the event-by-event basis. However, 
we immediately face the difficulty how to quantitatively measure the 
fluctuations in the events of very different multiplicity. The problem 
appears to be of more general nature. 

There are several interesting proposals to use fluctuations as a potential 
source of valuable information on the collision dynamics. If the hadronic 
system produced in the collision is in the thermodynamical equilibrium, 
the temperature and multiplicity fluctuations have been argued to determine,
respectively, 
the heat capacity \cite{Sto95,Shu98} and compressibility \cite{Mro98a} of 
the hadronic matter at freeze-out. An extensive discussion of the equilibrium 
fluctuations can be found in \cite{Ste99}. In the experimental realization of 
such ideas one has to disentangle however the `dynamical' fluctuations of 
interest from the `trivial' geometrical ones due to the impact parameter 
variation. The latter fluctuations are very sizable and dominate the 
fluctuations of all extensive event characteristics such as multiplicity 
or transverse energy. The variation of the impact parameter can also influence 
the fluctuations of the intensive quantities e.g. the temperature.

A specific solution to the problem was given in our paper \cite{Gaz92},
where we introduced the measure of fluctuations or correlations which 
has been later called $\Phi$. It is constructed in such a way that $\Phi$ 
is exactly the same for nucleon-nucleon (N--N) and  nucelus-nucelus 
(A--A) collisions if  the A--A collision is a simple superposition of 
N--N interactions. On the other hand, $\Phi$ equals zero when the 
correlations are absent in the collision final state. The method proposed 
in \cite{Gaz92} has been recently applied to the NA49 experimental data. 
The fluctuations of transverse momentum found in the central Pb--Pb 
collisions at 158 GeV per nucleon have appeared to be surprisingly small
\cite{Rol98,App99}. It has been also  claimed \cite{App99} that the 
correlations, which are of short range in the momentum space, are responsible 
for the nonzero positive value of  $\Phi_{p_T}$ being observed. Our 
calculations of $\Phi_{p_T}$ in the equilibrium ideal gas show 
\cite{Mro98b} that $\Phi_{p_T}$ is positive for bosons, negative 
for fermions and zero for classical particles.  When the hadronic system 
at freeze-out is identified with the pion gas, the calculated $\Phi_{p_T}$ 
slightly overestimates the experimental value \cite{App99} but the inclusion 
of the pions which come from the resonance decays removes the discrepency. 
An interesting analysis of the $p_T-$fluctuations within the so-called 
non-extensive statistics is given in \cite{Alb99}.

The theoretical analysis of the result \cite{Rol98,App99} has provided 
a new insight into the collision dynamics. It has been argued \cite{Ble98} 
within the UrQMD model that the secondary scatterings are responsible for the 
dramatic correlation loss in the central collisions of heavy-ions. This 
conclusion however seems to contradict the results of the analysis \cite{Liu98}
where the LUCIAE event generator has been used and the rescatterings 
are shown to reduce insignificantly the $p_T-$correlations measured by $\Phi$. 
While the effect of  the secondary interactions needs to be clarified, the 
smallness of the fluctuations observed in the central heavy-ion collisions 
\cite{Rol98,App99} is a very restrictive test of the collision models. 
The so-called random walk model is ruled out because it gives much stronger 
correlations in A--A than N--N case \cite{Gaz99}. The same holds \cite{Liu98} 
for the LUCIAE event generator when the jet production and/or the string 
clustering is taken into account even at a rather moderate rate. On the other 
hand, the quark-gluon string model seems to pass the test successfully 
\cite{Cap99}.

As argued in \cite{Gaz99}, the measure $\Phi$ can be also applied
to study the fluctuations of  chemical composition of the hadronic
system produced in the nuclear collisions. The chemical fluctuations
seem to be even more interesting than those of the kinematical 
variables such as $p_T$. The final state momentum distribution 
of hadrons characterizes the system at the moment of freeze-out, 
while the system chemical composition is fixed at the earlier evolution 
stage - the chemical freeze-out when the secondary {\it inelastic} 
interactions are no longer effective. The total strangeness yield 
presumably saturates even earlier and the subsequent interactions
are mostly responsible for the strangeness redistribution among
hadron species. 

The NA49 Collaboration plans to study the chemical fluctuations
in heavy-ion collisions at CERN SPS \cite{NA498}. Since the 
$\Phi-$measure will be used in these studies, it is desirable
to better understand the properties of $\Phi$ when applied to the chemical
fluctuations. This is the aim of our note. At the beginning we express 
$\Phi$ through the commonly used moments of the multiplicity 
distribution and analyze the result within several models of the 
distribution. Then, we compute the $\Phi-$measure for case of the 
two-component ideal quantum gas in equilibrium. The system 
of kaons and pions is discussed in detail. In particular, the
role of resonances is analysed.

Let us first introduce the measure $\Phi$ which describes the correlations 
(or fluctuations) of a single particle variable $x$ such as the particle 
energy or transverse momentum. As observed in \cite{Gaz99}, $x$ can 
also characterize the particle sort. Then, $x = 1$ if the particle is 
of the sort of interest, say the particle is strange, and $x= 0$ if the 
particle is {\it not} of this sort, it is a non-strange particle. We define 
the single-particle variable $z \buildrel \rm def \over = x - \overline{x}$ 
with the overline denoting averaging over a single particle inclusive 
distribution. In the case of the chemical fluctuations, $\overline{x}$ 
is the probability (averaged over events and particles) that a produced 
particle is of the sort of interest, say it is strange. One easily observes 
that $\overline{z} = 0$. We now introduce the event variable $Z$, which is 
a multiparticle analog of $z$, defined as 
$Z \buildrel \rm def \over = \sum_{i=1}^{N}(x_i - \overline{x})$, 
where the summation runs over particles from a given event.
By construction $\langle Z \rangle = 0$, where $\langle ... \rangle$ 
represents averaging over events. Finally, the $\Phi-$measure is defined 
in the following way
\begin{equation}\label{phi}
\Phi \buildrel \rm def \over = 
\sqrt{\langle Z^2 \rangle \over \langle N \rangle} -
\sqrt{\overline{z^2}} \;.
\end{equation}

We compute $\Phi$ for the system of particles of two sorts, $a$ and 
$b$, e.g. strange and non-strange hadrons. $x_i = 1$ when $i-$th 
particle is of the $a$ type and $x_i = 0$ otherwise. The inclusive 
average of $x$ and $x^2$ read
$$
\overline{x} = \sum_{x=0,1}xP_x = P_1 \;, 
\;\;\;\;\;\;\;\;\;\;\;\;\;\;\;\;\;
\overline{x^2} = \sum_{x=0,1}x^2P_x = P_1 \;,
$$
where $P_1$ is the probability (averaged over particles and events) 
that a produced particle is of the $a$ sort. Thus,
$$
P_1 = { \langle N_a \rangle \over \langle N_a \rangle +
\langle N_b \rangle } \;,
$$
with $N_a$ and $N_b$ being the numbers of particles $a$ and 
$b$, respectively, in a single event. One immediately finds 
that $\overline{z} = 0 $ while
\begin{eqnarray}\label{222}
\overline{z^2} = P_1 - P_1^2 = 
{ \langle N_a \rangle \langle N_b \rangle \over 
\langle N \rangle^2 } \;,
\end{eqnarray}
where $N = N_a + N_b$ is the multiplicity of all particles $a$ and $b$ 
in a single event.

Since the event variable $Z$ equals $N_a - \overline{x}N$, one gets
\begin{eqnarray*}
\langle Z \rangle &=& \langle N_a \rangle - \overline{x}
\langle N \rangle = 0 \;,\\[2mm]
\langle Z^2 \rangle &=& \langle N_a^2 \rangle 
- 2 \overline{x} \langle N_aN \rangle
+ \overline{x}^2 \langle N^2 \rangle \;.
\end{eqnarray*}
The latter equation gives
$$
\langle Z^2 \rangle \langle N \rangle^2 = 
\langle N_b \rangle^2 \langle N_a^2 \rangle
+\langle N_a \rangle^2 \langle N_b^2 \rangle 
- 2 \,\langle N_a \rangle \langle N_b \rangle 
\langle N_a N_b \rangle \;,
$$
which can be rewritten as
\begin{eqnarray}\label{111}
{\langle Z^2 \rangle \over \langle N \rangle} =
{\langle N_b \rangle^2 \over \langle N \rangle^3}
\big(\langle N_a^2 \rangle - \langle N_a \rangle^2 \big)
&+&{\langle N_a \rangle^2 \over \langle N \rangle^3} 
\big(\langle N_b^2 \rangle - \langle N_b \rangle^2 \big) 
\\ [3mm] \nonumber
&-& 2 \,{\langle N_a \rangle \langle N_b \rangle \over \langle N \rangle^3}
\big(\langle N_a N_b \rangle - \langle N_a \rangle \langle N_b \rangle\big)
\;.
\end{eqnarray}
The fluctuation measure $\Phi$ is completely determined by 
eqs. (\ref{222}, \ref{111}). So, let us consider its properties 
within three simple models of the multiplicity distribution.

{\bf 1)} The distributions of particles $a$ and $b$ are poissonian 
and independent from each other i.e. 
\begin{eqnarray}\label{zero}
\langle N_i^2 \rangle - \langle N_i \rangle^2 &=& \langle N_i \rangle 
\;,
\\[3mm] \nonumber
\langle N_a N_b \rangle &=& \langle N_a \rangle \langle N_b \rangle \;,
\end{eqnarray}
where $i=a,b$. One easily notices that $\Phi=0$ in this case. 

{\bf 2)} The particles $a$ and $b$ are assumed to be correlated in such 
a way that there are {\it no} chemical fluctuations in the system. The 
event chemical composition, which is fully characterized (for a two 
component system) by the ratio $N_a/N$, is assumed to be strictly 
independent of the event multiplicity. Then, $N_a = \alpha \, N$ and 
$N_b = (1 - \alpha) \, N$ with $\alpha$ being a constant smaller than 
unity. Since $N_a$, $N_b$ and $N$ are the integer numbers, $\alpha$ 
has to be a rational fraction. Then, we have
$$
\langle N_a \rangle = \alpha \,\langle N \rangle \;,
\;\;\;\;\;\;\;\;\;\;\;
\langle N_a \rangle = ( 1- \alpha ) \,\langle N \rangle \;, 
$$
$$
\langle N_a^2 \rangle - \langle N_a \rangle ^2 = 
\alpha^2 \big(\langle N^2 \rangle - \langle N \rangle^2 \big) \;,  
\;\;\;\;\;\;\;\;\;\;\;
\langle N_b^2 \rangle - \langle N_b \rangle ^2 = 
(1-\alpha)^2 \big(\langle N^2 \rangle - \langle N \rangle^2 \big) \;, 
$$
$$
\langle N_a N_b \rangle  - \langle N_a \rangle \langle N_b \rangle  = 
\alpha (1-\alpha)\, \big(\langle N^2 \rangle - \langle N \rangle^2 \big) \;.
$$
Consequently, $\langle Z^2 \rangle = 0$ and 
\begin{equation}\label{zero-chem}
\Phi =  - \sqrt{{ \langle N_a \rangle \langle N_b \rangle \over 
\langle N \rangle^2 }} = - \sqrt{\alpha (1- \alpha)} \;.
\end{equation}
One sees that the $\Phi-$measure is negative (but larger than $-1/2$) 
when the chemical fluctuations vanish in the system. 

{\bf 3)} The particles $a$ and $b$ are identified with the positively and, 
respectively, negatively charged hadrons. Then, the charge conservation
leads to the strict correlation of the particle numbers:
$$
N_+ - N_- = Q \;,
$$
where $Q$ is the electric charge of the system. In this case we have
$$
\langle N_+ \rangle = \langle N_- \rangle + Q \;, 
$$
$$
\langle N_+^2 \rangle - \langle N_+ \rangle ^2 = 
\langle N_-^2 \rangle - \langle N_- \rangle^2  \;, 
$$
$$
\langle N_+ N_- \rangle -\langle N_+ \rangle \langle N_- \rangle = 
\langle N_-^2 \rangle - \langle N_- \rangle^2  \;.
$$
Therefore,
\begin{eqnarray*}
\overline{z^2} = {\big( \langle N_- \rangle + Q \big) 
\langle N_- \rangle \over \langle N \rangle^2 } \;,\\[3mm]
{\langle Z^2 \rangle \over \langle N \rangle} = 
{Q^2 \over \langle N \rangle^3}
\big( \langle N_-^2 \rangle - \langle N_- \rangle^2 \big) \;.
\end{eqnarray*}
When $Q = 0$ we reproduce the result (\ref{zero-chem}) corresponding 
to $\alpha = 1/2$.

After the illustrative examples let us compute $\Phi$ for the equilibrium 
gas which is a mixture of the particles $a$ and $b$. Then, 
\begin{eqnarray}\label{meanN1}
\langle N_i \rangle &=& \lambda_i {\partial \over  \partial \lambda_i}
{\rm ln}\,\Xi(V,T,\lambda_a,\lambda_b) \;, 
\\[3mm] \nonumber
\langle N_a N_b \rangle - \langle N_a \rangle \langle  N_b \rangle 
&=& \lambda_a \lambda_b 
{\partial^2 \over  \partial \lambda_b \partial \lambda_a } 
{\rm ln}\,\Xi(V,T,\lambda_a,\lambda_b)  \;,
\\[3mm] \nonumber
\langle N_i^2 \rangle - \langle N_i \rangle^2 &=& 
\bigg(\lambda_i {\partial \over  \partial \lambda_i}\bigg)^2\;
{\rm ln}\,\Xi(V,T,\lambda_a,\lambda_b) \;, 
\end{eqnarray}
where $\Xi(V,T,\lambda_a,\lambda_b)$ is the grand canonical partition 
function with $V$, $T$ and $\lambda_i$ denoting, respectively, the system 
volume, temperature and the fugacity which is related to the chemical 
potential $\mu_i$ as $\lambda_i = e^{\beta \mu_i}$ with $\beta \equiv T^{-1}$.

When the gas of interest is the mixture of the two ideal quantum gases,
the partition function is \cite{Hua63}
\begin{equation}\label{partition}
{\rm ln}\,\Xi(V,T,\lambda_a,\lambda_b) = 
\pm g_a\,V  \int{d^3p \over (2\pi )^3} \;
{\rm ln}\big[1 \pm \lambda_a\,e^{-\beta E_a} \big] 
\pm g_b\,V  \int{d^3p \over (2\pi )^3} \;
{\rm ln}\big[1 \pm \lambda_b\,e^{-\beta E_b} \big] \;,
\end{equation}
where $g_i$ denotes
the number of the particle internal degrees of freedom; 
$E_i \equiv \sqrt{m_i^2 + {\bf p}^2}$ is the particle energy with $m_i$ and 
${\bf p}$ being its mass and momentum; the upper sign is for fermions 
while the lower one for bosons. 

Substituting the ideal gas partition function (\ref{partition}) into 
eqs. (\ref{meanN1}), one easilly finds 
\begin{eqnarray}\label{meanN2}
\langle N_i \rangle &=& 
g_iV \,\int{d^3p \over (2\pi )^3} \;
{1 \over \lambda_i^{-1}e^{\beta E_i} \pm 1}\;, 
\\[3mm] \nonumber
\langle N_a N_b \rangle  &=& 
\langle N_a \rangle \langle N_b \rangle \;,
\\[3mm] \nonumber
\langle N_i^2 \rangle - \langle N_i \rangle^2 &=& 
g_iV \,\int{d^3p \over (2\pi )^3} \;
{\lambda_i^{-1}e^{\beta E_i}
\over (\lambda_i^{-1}e^{\beta E_i} \pm 1)^2}
\;, 
\end{eqnarray}
where, as presviously, the index $i$ labels the particles of the type
$a$ or $b$. It is worth noting that the system volume $V$ which enters 
eqs.~(\ref{meanN2}) cancels out in the final expression of $\Phi$. 
Therefore, the measure $\Phi$ is, as expected, an intensive quantity. 
One observes in eqs.~(\ref{meanN2}) that
$$
\langle N_i^2 \rangle - \langle N_i \rangle ^2  <  \langle N_i \rangle 
$$
for fermions,
$$
\langle N_i^2 \rangle - \langle N_i \rangle ^2  >  \langle N_i \rangle 
$$
for bosons, and
$$
\langle N_i^2 \rangle - \langle N_i \rangle ^2  =  \langle N_i \rangle 
$$
in the classical limit where $\lambda_i^{-1} \gg 1$. Therefore, one finds
from eqs. (\ref{222}, \ref{111}) that
$$
{\langle Z^2 \rangle \over \langle N \rangle } < \overline{z^2}
\;\;\;\;\;\;\; {\rm and} \;\;\;\;\;\;\; \Phi < 0
$$
when the particles $a$ and $b$ are fermions,
$$
{\langle Z^2 \rangle \over \langle N \rangle }> \overline{z^2}
\;\;\;\;\;\;\; {\rm and} \;\;\;\;\;\;\; \Phi > 0
$$
when the particles $a$ and $b$ are bosons, and
$$
{\langle Z^2 \rangle \over \langle N \rangle }= \overline{z^2}
\;\;\;\;\;\;\; {\rm and} \;\;\;\;\;\;\; \Phi = 0
$$
when the particles of  both types can be treated as classical. 
If the particles $a$ and $b$ are of different statistics, the sign 
of $\Phi$ is determined by the sign of the expression 
\begin{eqnarray*}
\langle N_a \rangle^2  
\big( \langle N_b^2 \rangle - \langle N_b \rangle ^2 
- \langle N_b \rangle \big) 
+ \langle N_b \rangle^2  
\big( \langle N_a^2 \rangle - \langle N_a \rangle ^2 
- \langle N_a \rangle \big) \;,
\end{eqnarray*}
which is either positive or negative depending of the particle masses, 
their chemical potentials, and the numbers of the internal degrees of freedom.

If all particles are massless and their chemical potentials 
vanish, the calculations can be performed analytically. In this 
case eqs. (\ref{meanN2}) give
\begin{eqnarray*}
\langle N_i \rangle &=& 
{g_i\zeta(3) \over \pi^2} {3/4 \choose 1}\, V T^3
\cong g_i {0.09 \choose 0.12} \, V T^3
\;, \\[3mm] 
\langle N_i^2 \rangle - \langle N_i \rangle^2 &=& 
{g_i \over 6} {1/2 \choose 1}\, V T^3
\cong g_i {0.08 \choose 0.17} \, V T^3 \;, 
\end{eqnarray*}
where $\zeta(x)$ is the Riemann zeta function ($\zeta(3) \cong 1.202$); 
as previously the upper case is for fermions and the lower one for bosons.
If the particles $a$ and $b$ are both fermions or both bosons, 
eqs. (\ref{222}, \ref{111}) get the form
\begin{eqnarray}\label{1and2}
\overline{z^2} &=& {g_a g_b \over (g_a + g_b)^2} \;,
\\[3mm] \nonumber
{\langle Z^2 \rangle \over \langle N \rangle }&=&
{\pi^2 \over 6 \zeta(3)}\; { g_a g_b \over (g_a + g_b)^2}
\; {2/3 \choose 1} \;,
\end{eqnarray}
and consequently 
$$
\Phi \cong {- 0.045 \choose \;\;\; 0.170} 
{\sqrt{g_a g_b} \over g_a + g_b} \;.
$$

Let us now consider the fluctuations in the system of pions and kaons.
To be specific, the particles $a$ are identified with $\pi^-$ while the 
particles $b$ with $K^+$ or $K^-$. As we will see, the fluctuations in 
the $\pi^-K^+$ system can be very different from those in $\pi^-K^-$ one. 
The systems $\pi^+K^+$ and $\pi^+K^-$, which are not discussed here, are 
analogous to, respectively, $\pi^-K^-$ and $\pi^-K^+$.  At first we treat 
the pions and kaons as a mixture of the ideal gases of $\pi$ and $K$. Since 
the pions and kaons are of a given charge (plus or minus) $g_{\pi} = g_K = 1$. 
The masses are taken, respectively, 140 and 494 MeV. $\Phi$ as a function of
temperature has been computed numerically from the formulas 
(\ref{phi},\ref{222},\ref{111}) combined with (\ref{meanN2}). The results, 
which are obviously the same for the $\pi^-K^+$ and $\pi^-K^-$ systems, 
are shown with the dashed lines in Figs. 1-4. The calculations have been 
performed for several values of the chemical potentials of pions and kaons. 
In the case of pions, $\mu_{\pi} = 0$ when the system is in the chemical 
equilibrium. (We obviously neglect here a tiny effect of the electric charge 
conservation.)  The finite value of $\mu_K$ appears even in the equilibrium 
system of  zero net strangeness due to the simultaneous baryon and strangeness 
conservation. For example, the estimated value of $\mu_K$ for the strange 
(not antistrange) mesons is 38 MeV at $T = 160$ MeV for the equilibrium
hadronic system produced in heavy-ion collisions at CERN SPS \cite{Bra96}.

In Figs. 1 and 2 we observe a dramatic increase of $\Phi$ with the 
temperature. $\Phi$ also grows with $\mu_K$ (at $\mu_{\pi}=0$) while 
the dependence on $\mu_{\pi}$ changes with the temperature. Below 
$T \cong 120$ MeV $\Phi$ is a decreasing function of $\mu_{\pi}$ but 
above this temperature $\Phi$ grows with $\mu_{\pi}$ ($\mu_K$ is fixed 
and equals zero). Such a behaviour can be easily understood. $\Phi$ can 
be approximated as
\begin{equation}\label{phi-app}
\Phi \cong { \sqrt{ \langle N_K^2 \rangle - \langle N_K \rangle^2} 
-\sqrt{\langle N_K \rangle} \over \sqrt{\langle N_{\pi} \rangle}} \;,
\end{equation}
for $\langle N_{\pi} \rangle \gg \langle N_K \rangle$ which holds for 
sufficiently low temperatures. (We also assume here that 
$\langle N_K^2 \rangle - \langle N_K \rangle^2$ is {\it not} larger than 
$\langle N_{\pi}^2 \rangle - \langle N_{\pi} \rangle^2$.) The growth of 
$\mu_K$ leads to the increase of the numerator of  the expression 
(\ref{phi-app}) while the growth of $\mu_{\pi}$ enlarges the 
denominator. At higher temperatures the numbers of pions and 
kaons are comparable to each other and the pion dispersion 
$\langle N_{\pi}^2 \rangle - \langle N_{\pi} \rangle^2$ provides
a significant contribution to $\Phi$. Then, $\Phi$ grows with 
$\mu_{\pi}$.

It is a far going idealization to model a fireball at freeze-out 
as an ideal gas of pions and kaons. A substantial fraction of the 
final state particles come from the hadron resonances. We take them
into account in the following way. Since the resonances are relatively
heavy, their phase-space density is rather low. Consequently, the 
resonances can be treated as classical particles with the poissonian
multiplicity distribution \cite{Ste99}. Then, one easily shows that
\begin{eqnarray}\label{reso}
\langle N_i \rangle &=& \langle N_i^\prime \rangle
+ \sum_r b_r \langle N_r \rangle \;, \;\;\;\;\;\;\;\;i = \pi, K \\[2mm]
\langle N_i^2 \rangle - \langle N_i \rangle^2 &=&
\langle {N_i^\prime}^2 \rangle - \langle N_i^\prime \rangle^2
+ \sum_r b_r \langle N_r \rangle \;, \nonumber \\ [2mm]
\langle N_{\pi}N_K \rangle - 
\langle N_{\pi}\rangle \langle N_K \rangle  &=&
\langle N_{\pi}^{\prime} N_K^{\prime} \rangle 
- \langle N_{\pi}^{\prime} \rangle \langle N_K^{\prime} \rangle 
+ \sum_r b_r^* \langle N_r^* \rangle \;, \nonumber
\end{eqnarray}
where $N_i^\prime$ is the number of `direct' pions or kaons which are 
described by the formulas (\ref{meanN2}); the summation runs over the 
resonances which decay into pions or kaons; $N_r$ is the number of 
resonances of type $r$ and $b_r$ is the branching ratio of the resonance 
decay into the pion or kaon channel. The resonances are assumed to decay 
into no more than one pion or one kaon of interest. We denote with $*$ 
the resonances, such as $K^*$, which decay into the pion-kaon pair under 
study. 

In the actual calculations we have taken into account the lightest resonances: 
$\rho(770)$, $\omega(782)$ and $K^*(892)$. Now, an important difference 
between the correlations in the $\pi^-K^-$ and $\pi^-K^+$ system appears. 
The decays of $\bar K^{*0}$ into $K^+\pi^-$ produce the correlation in the 
$\pi^-K^+$ system. Analogous correlation in the $\pi^-K^-$ system is absent. 
$\Phi$ as a function of temperature has again been computed 
numerically from the formulas (\ref{phi},\ref{222},\ref{111}) combined 
with eqs.~(\ref{meanN2}) which are now supplemented with eqs.~(\ref{reso}). 
The results are shown with the solid lines in Figs. 1 and 2 for the 
$\pi^-K^-$ system and in Fig. 3 and 4 for the $\pi^-K^+$ 
one. The calculations have been again performed for several values of 
the chemical potentials of pions and kaons. The chemical potential
of $\rho$ and $\omega$ has been taken to be equal to $\mu_{\pi}$ while
that of $K^*$ equals $\mu_K$. One sees that in the case of $\pi^-K^-$ 
correlations the presence of resonances does not change the results 
qualitatively although the value of $\Phi$ is significantly reduced. 
The case of  $\pi^-K^+$ is changed dramatically due to the resonances. 
The role of the term corresponding to 
$\langle N_{\pi}N_K \rangle - \langle N_{\pi}\rangle \langle N_K \rangle $ 
appears to be so important that $\Phi$ becomes negative for sufficiently 
large temperatures. It is somewhat surprising that $\Phi$ from Fig. 4 changes 
its sign at the temperature of about $T= 110$ MeV which is approximately 
{\it independent} of $\mu_K$. Such a behaviour can be understood as follows. 
Since $\langle N_{\pi}\rangle > \langle N_K \rangle$ in the domain of the 
parameter values of interest, we expand the expressions (\ref{111}) and 
(\ref{222}) in powers of $\langle N_K \rangle / \langle N_{\pi}\rangle $. One 
observes that the first power terms of (\ref{111}) and (\ref{222}) cancel out 
each other. Therefore, $\Phi=0$ when the second power terms of (\ref{111}) 
and (\ref{222}) are equal to each other. Taking into account
that the kaons are approximately classical and consequently $\Phi$
depends on $\mu_K$ roughly as $e^{\beta \mu_K}$, one indeed finds 
that the position of $\Phi = 0$ is approximately independent of $\mu_K$.

At the end we take an effort to estimate $\Phi$  from the existing 
experimental data \cite{Bar74} which appear to be rather scarce. Specifically, 
we consider the system of $K^0_s$ and negative hadrons produced in $pp$ 
interactions at the energy 205 GeV, which is close to the currently available 
energies of  heavy-ion collisions at CERN SPS. This case is expected to be 
similar to the $\pi^-K^+$ system discussed above. One finds in \cite{Bar74} 
that:
$$
\langle N_- \rangle = 2.84 \;,
\;\;\;\;\;\;\;\;\;\;
\langle N_-^2 \rangle - \langle N_- \rangle^2 = 3.63
$$ 
$$
\langle N_K \rangle =  0.18 \;,
\;\;\;\;\;\;\;\;\;\;
\langle N_-N_K \rangle - \langle N_-\rangle \langle N_K \rangle = 0.078 \;.
$$
Unfortunately, there are no data on 
$\langle N_K^2 \rangle - \langle N_K \rangle^2$ which gives 
a dominant contribution to $\Phi$. If  the multiplicity distribution of kaons 
is poissonian i.e. 
$\langle N_K^2 \rangle - \langle N_K \rangle^2 = \langle N_K \rangle$, 
we get $\Phi = - 0.004$. The dispersion of the negative hadron multiplicity 
distribution is known to follow the so-called Wr\'oblewski formula 
\cite{Wro73}. Applying the formula to the kaons we have 
$\langle N_K^2 \rangle - \langle N_K \rangle ^2 = 
\big( 0.58 \langle N_K \rangle + 0.29 \big)^2 = 0.16$. In this case the
kaon multiplicity appears to be even narrower than the poissonian one and 
$\Phi =- 0.02$.  The estimate of $\Phi = 0.007$ given in \cite{Gaz99} exceeds
the two our numbers because the kaon mutiplicity distribution, which is used 
in \cite{Gaz99}, is (after averaging over the negative hadron 
multiplicity) broader than the poissonian one. We conclude that the 
existing data give a rather poor information on $\Phi$ in $pp$ collisions.

We summarize our study as follows. The $\Phi-$measure seems to be a
useful tool to study the chemical fluctuations in heavy-ion collisions.
If the particles of different species are produced independently from 
each other and the multiplicity distributions are poissonian, $\Phi$
is exactly zero. When the particles are produced in such a way that
there are no chemical fluctuations  (particle ratios are fixed), $\Phi$
is negative but larger than $-1/2$. If the nucleus-nucleus collision is 
a simple superposition of N--N interactions the value $\Phi$ is strictly 
independent of the collision centrality. The same happens when the 
hadronic system produced in the nucleus-nucleus collisions achieves 
the equilibrium with the temperature and chemical potentials being 
independent of the impact parameter. The thermal model, which seems 
to be successful in describing the average multiplicities of different 
particle species, gives a definite prediction of  $\Phi$, which is positive
for bosons and negative for fermions. The correlations in the system of
pions and kaons have been considered in detail. The estimate of $\Phi$ 
for the $\pi^-K^-$ system is rather reliable while the prediction concerning 
the $\pi^-K^+$ correlations is sensitive to the details of the model. Since 
the experimental value of  $\Phi$ in $pp$ interactions can be hardly extracted
from the existing data, the fluctuation measurements of nuclear collisions 
should start with the nucleon-nucleon case.

\vspace{1cm}

I am very indebted to ECT* in Trento where the idea to analyze the 
equilibrium chemical fluctuations was born. Numerous fruitful discussions 
with Marek Ga\' zdzicki, who initiated this study, are also gratefully 
acknowledged.  


\newpage
\vspace{1cm}
\begin{center}
{\bf Figure Captions}
\end{center}
\vspace{0.3cm}

\noindent
{\bf Fig. 1.} 
$\Phi-$measure of the $\pi^-K^-$ correlations as a function 
of temperature for three values of the pion chemical potential. The kaon 
chemical potential vanishes. The resonances are either neglected (dashed
lines) or taken into account (solid lines). The most upper dashed line on the 
right hand side of the figure corresponds to $\mu_{\pi} = 100$ MeV, the 
central one to $\mu_{\pi} = 0$, and the lowest line to $\mu_{\pi} = -100$ 
MeV. At sufficiently small temperatures the respective dashed and solid
lines coincide. 

\vspace{0.5cm}

\noindent
{\bf Fig. 2.} 
$\Phi-$measure of the $\pi^-K^-$ correlations as a function 
of temperature for three values of the kaon chemical potential. The pion
chemical potential vanishes. The resonances are either neglected (dashed
lines) or taken into account (solid lines). The most upper dashed line 
corresponds to $\mu_K = 100$ MeV, the central one to $\mu_K = 0$, and 
the lowest line to $\mu_K= -100$ MeV. At sufficiently small temperatures 
the respective dashed and solid lines coincide. 

\vspace{0.5cm}

\noindent
{\bf Fig. 3.} 
The absolute value of $\Phi-$measure of the $\pi^-K^+$ correlations 
as a function of temperature for three values of the pion chemical potential. 
The kaon chemical potential vanishes. The resonances are either neglected (dashed
lines) or taken into account (solid lines). The most upper dashed line on the 
right hand side of the figure corresponds to $\mu_{\pi} = 100$ MeV, the 
central one to $\mu_{\pi} = 0$, and the lowest line to $\mu_{\pi} = -100$ 
MeV. At sufficiently small temperatures the respective dashed and solid
lines coincide. 

\vspace{0.5cm}

\noindent
{\bf Fig. 4.} 
The absolute value of  $\Phi-$measure of the $\pi^-K^+$ correlations 
as a function of temperature for three values of the kaon chemical potential. 
The pion chemical potential vanishes. The resonances are either neglected 
(dashed lines) or taken into account (solid lines). The most upper dashed and 
solid lines correspond to $\mu_K = 100$ MeV, the central ones to $\mu_K = 0$, 
and the lowest lines to $\mu_K= -100$ MeV. At sufficiently small temperatures 
the respective dashed and solid lines coincide.


\begin{thebibliography}{99}

\bibitem{Koc86} P.~Koch, B.~M\"uller and J.~Rafelski, 
Phys. Rep. {\bf 142} (1986) 167.

\bibitem{Gaz96} M.~Ga\' zdzicki and D.~R\"orich, 
Z. Phys. {\bf C71} (1996) 55.

\bibitem{Ody98} G.~J.~Odyniec, Nucl. Phys. {\bf A638} (1998) 135c.

\bibitem{Raf96} J.~Rafelski, J.~Letessier and A.~Tounsi,
Acta Phys. Pol. {\bf B27} (1996) 1037.

\bibitem{Gaz97a} M.~Ga\' zdzicki, J. Phys. {\bf G23} (1997) 1881.

\bibitem{Bra96} P.~Braun-Munzinger et al., Phys. Lett. {\bf B365} (1996) 1.

\bibitem{Cap96} A.~Capella et al., Z. Phys. {\bf C70} (1996) 507.

\bibitem{Sor98} H.~Sorge, Nucl. Phys. {\bf  A630} (1998) 522c.

\bibitem{Sto95} L.~Stodolsky, Phys. Rev. Lett. {\bf 75} (1995) 1044.

\bibitem{Shu98} E.V.~Shuryak, Phys. Lett. {\bf B423} (1998) 9.

\bibitem{Mro98a} St.~Mr\'owczy\'nski, Phys. Lett. {\bf B430} (1998) 9.  

\bibitem{Ste99} M.~Stephanov, K.~Rajagopal and E.~Shuryak, hep-ph/9903292.

\bibitem{Gaz92} M.~Ga\' zdzicki and St.~Mr\' owczy\' nski, 
Z. Phys. {\bf C54} (1992) 127.

\bibitem{Rol98} G.~Roland and NA49 Collaboration, Nucl. Phys. 
{\bf A638} (1998) 91c.

\bibitem{App99} H.~Appelsh\"auser et al., hep-ex/9904014.

\bibitem{Mro98b} St.~Mr\'owczy\'nski, Phys. Lett. {\bf B439} (1998) 6.

\bibitem{Alb99} W.M.~Alberico, A.~Lavagno and P.~Quarati, nucl-th/9902070.

\bibitem{Ble98} M.~Bleicher et al., Phys. Lett. {\bf B435} (1998) 9.

\bibitem{Liu98} F.~Liu et al., hep-ph/9809320, Euro. Phys. J. {\bf C} 
in print.

\bibitem{Gaz99} M.~Ga\' zdzicki, A.~Leonidov and G.~Roland, 
Euro. Phys. J. {\bf C6} (1999) 365.

\bibitem{Cap99} A.~Capella, E.G.~Ferreiro and A.B.~Kaidalov, hep-ph/9903338.

\bibitem{Gaz99} M.~Ga\' zdzicki, Euro. Phys. J. {\bf C8} (1999)131.

\bibitem{NA498} NA49 Collaboration, {\it Status and Future Program 
of NA49 Experiment}, Addendum \#2 to proposal SPSLC/P264, CERN,
Geneva, 1998.

\bibitem{Hua63} K.~Huang, {\it Statistical Mechanics} 
(John Wiley, New York, 1963).

\bibitem{Bar74} S.~Barish et al., Phys. Rev. {\bf D9} (1974) 2689; 
K.~Jaeger et al., {\it ibid.} {\bf D9} (1975) 2405.

\bibitem{Wro73} A.~Wr\'oblewski, Acta Phys. Pol. {\bf B4} (1973) 857. 

\end{thebibliography}
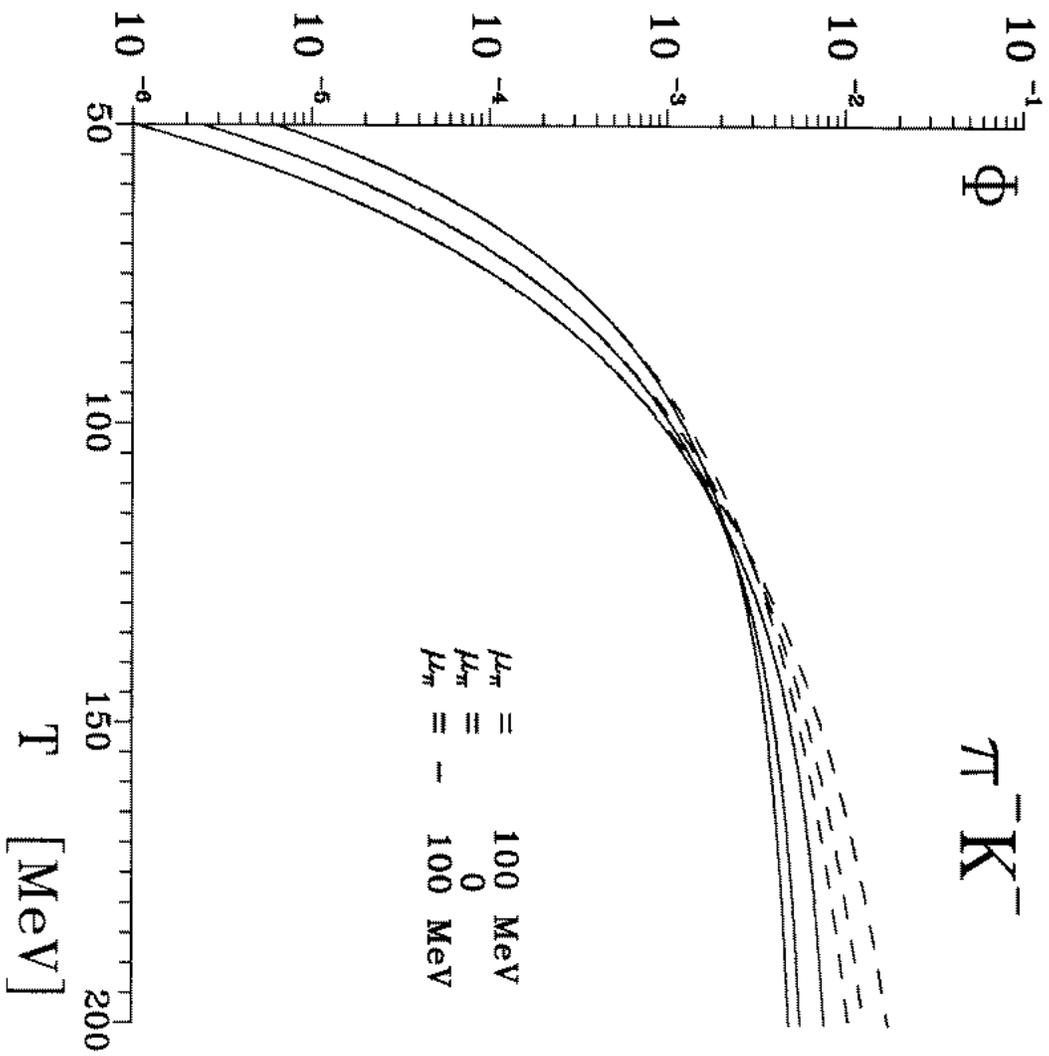
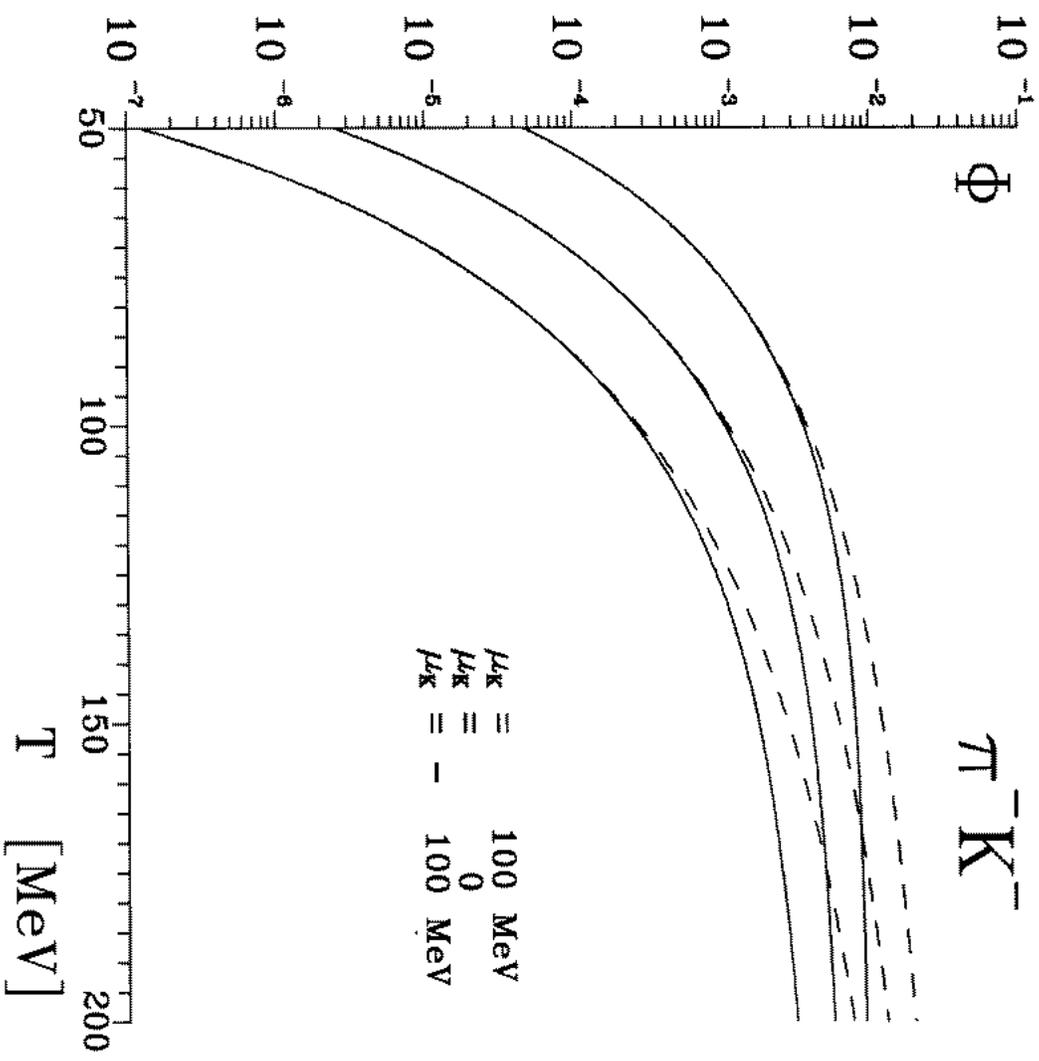
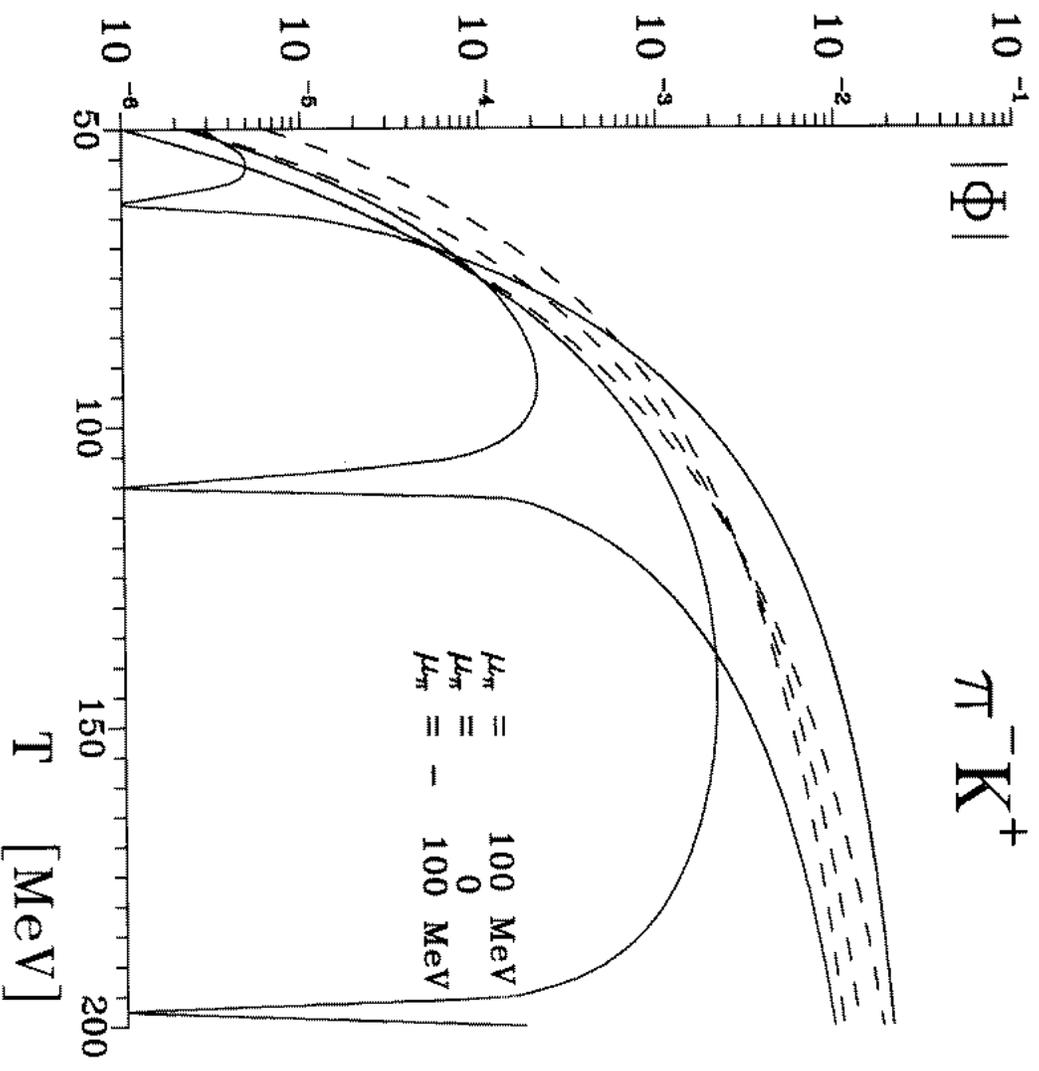
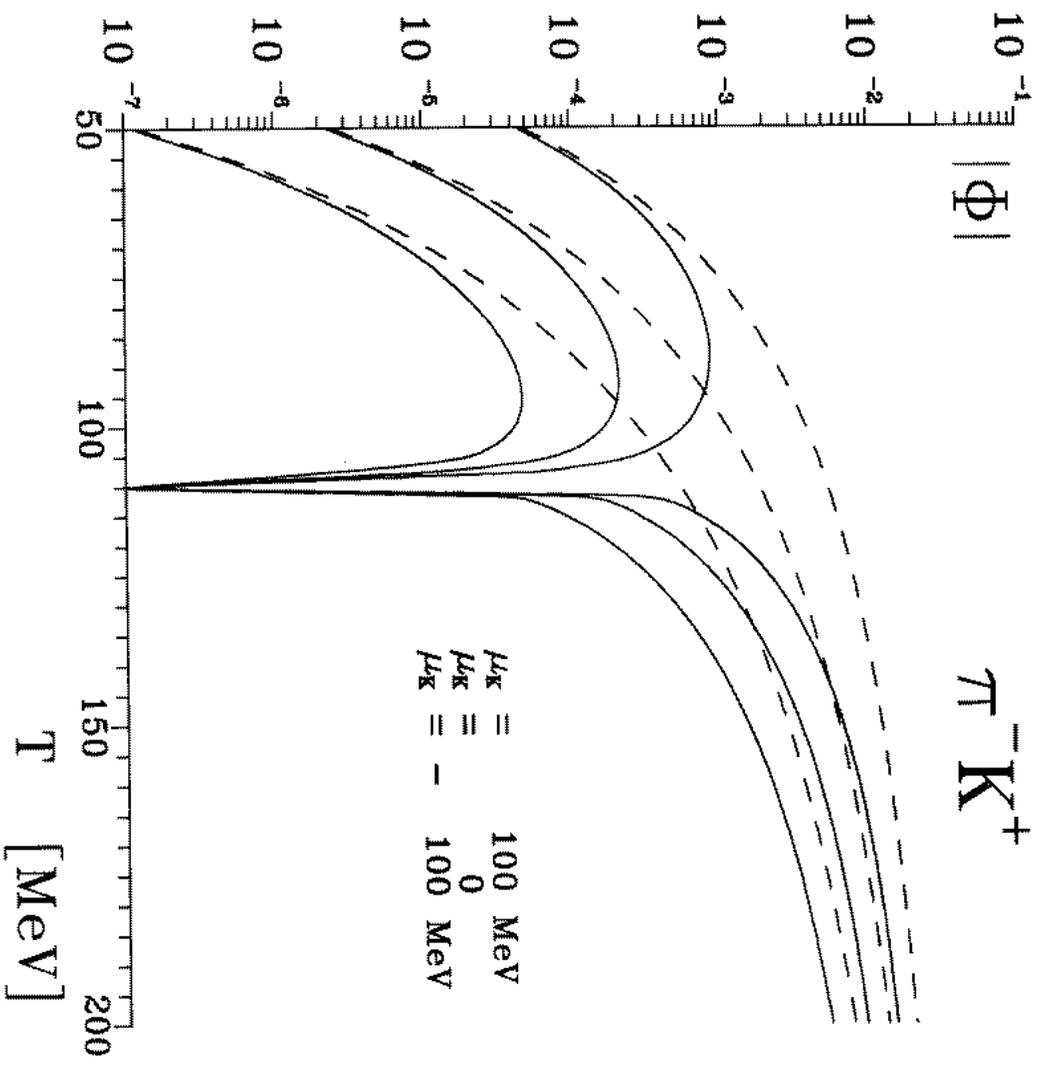
\end{document}